\documentclass[letterpaper,twocolumn,10pt]{article}
\usepackage[hyphens]{url}
\usepackage{breakurl}
\usepackage{usenix,epsfig,caption,subcaption}

\begin{document}

\title{\Large \bf The Zen of Graduate-level Programming}
\author{
{\rm Zubair Nabi}\thanks{Work done when the author was a part of
the faculty at Information Technology University, Pakistan.}\\
IBM Research, Dublin\\
zubairn@ie.ibm.com
}

\maketitle

\thispagestyle{empty}

\begin{abstract}
The ubiquity of technology in our daily lives and the economic stability of the
technology sector in recent years, especially in areas with a computer science
footing, has led to an increase in computer science enrollment in many parts of
the world. To keep up with this trend, the undergraduate computer science
curriculum has undergone many revisions, analysis, and discussion.
Unfortunately, the graduate level curriculum is lagging far behind in computer
science education literature and research. To remedy this, we present the
blueprint and execution of a graduate level course in programming, designed
specifically to cater to the needs of graduate students with a diverse
background both in CS and other fields. To this end, the course is divided into
two halves. In the first half, students are introduced to different programming
concepts, such as multi-paradigm programming, data structures, concurrency, and
security to bring them up to speed and provide a level playing field. In the
second half, all of these concepts are employed as building blocks to solve
real-world problems from data mining, natural language processing, computer
vision, and other fields. In addition, the paper also discusses in detail the
evaluation instruments employed for the course. Moreover, we also share
anecdotal information around student feedback, course design, and grading that
may be useful for others who wish to replicate our curriculum or sketch a
similar course.
\end{abstract}

\section{Introduction}
Programming is an integral part of computer science and many other scientific
fields. As a result, it has received widespread attention in computer science
education research as part of CS1 and
CS2~\cite{Turner:2005:TES,Howe:2004:CAC,Yarosh:2007:NDS,Caspersen:2009:SFP,Davies:2011:SCP}.
On the other hand, there is a dearth of any formal research on graduate level
programming courses. This is detrimental for the design of graduate computer
science programs which have been experiencing a consistent increase in
enrollment in the last few years~\cite{Zweben:2012:CDA}. The situation is
exacerbated by the fact that these programs also take in students who have no
formal background at the undergraduate level in computer science. In some cases,
students may not have written a single line of code prior to their enrollment.
Therefore, such students are forced to take introductory undergraduate level
courses to bridge that gap or to resort to self-learning. The former only
touches the surface in terms of skills required for graduate students while the
latter does not result in a comprehensive sweep of programming concepts.

The design of a graduate level programming course faces a number of challenges,
some of which it shares with its undergraduate counterparts while others are
unique to its specialized level. In the case of the former, challenges include
the widespread misconception that computer science equates to computer
programming. The focus on just programming in CS1 and CS2 results in students
mastering coding rather the theory behind it. In addition, students never think
like computer scientists because they are never introduced to how computer
scientists think about problems in their respective
domains~\cite{Turner:2005:TES}. Furthermore, there is a large disconnect between
how programming is taught in these courses and how it is actually employed in
the real-world~\cite{Howe:2004:CAC,Yarosh:2007:NDS}. Finally, the compositions
based teaching approach of programming enables students to learn bits and pieces
but they do not know how to put them together as an ensemble to solve a larger
problem~\cite{Caspersen:2009:SFP}. These challenges are exacerbated in graduate
programs due to the diverse background of students. Students with a non-CS
background think of programming as the most difficult part of the
curriculum~\cite{Lunney:2003:JLF}. Moreover, graduate students have a preference
for programming languages that they can directly use in the industry and improve
their chances of future employment. Finally, due to non-progression, there is a
wide gap between the end of the undergrad and the beginning of graduate studies
for many students~\cite{Lunney:2003:JLF}.

To remedy this, some institutions have started offering specialized foundational
courses to help students in transitioning from a non computer science background
into a graduate degree in CS. Interestingly, having these courses as part of the
graduate curriculum has increased enrollment in some
cases~\cite{Kongmunvattana:2012:AFC}. In addition, it can whet the appetite of
PhD students and prepare them to tackle domain specific challenges as
programming spans multiple disciplines~\cite{Arden:1969:RPP}. Furthermore,
students who have some background in the industry, can use existing knowledge to
enhance their graduate learning experience~\cite{Iyengar:1978:ECP}.

The Information Technology University in Pakistan offered a graduate level
course in programming in Spring 2013. This paper presents the design and
execution of the course, dubbed Advanced Real-world Programming
(ARWP)\footnote{The entire material for the course, including lectures,
assignments, quizzes, lab exercises, and projects is available online:
\url{https://sites.google.com/site/arwpspring13/}}. To the best of our
knowledge, this is the first time a graduate level programming course has been
presented in such detail. The overarching goal of the course is to motivate
graduate students to start thinking like computer scientists~\cite{elkner:think}
inspired by 6.00x at MIT~\cite{6.00x}. In a similar vein, Python is the language
of choice for the course due to its clean syntax, multi-paradigm support, large
suite of libraries, and wide
applicability~\cite{Davies:2011:SCP,Dodds:2008:EBC,Blank:2003:PPV}. The course
is divided into two parts, in the first half, Python is employed as a vehicle to
illustrate basic programming constructs, data structures, concurrency, different
programming paradigms, networking, and user interface design. In the second half
of the course, these building blocks are used to solve real-world problems from
machine learning, NLP, graph theory, and computer vision, to name a few. All
assignments and projects in the course are derived from real-world use-cases and
applications~\cite{Stevenson:2006:DRP} and are self-contained. Taking this one
step further, students also perform code reviews of each other's projects. In
addition, the weekly lab component of the course enables students to enhance the
learning experience by getting their hands dirty. Finally, the course also
incorporates programming for alternative environments such as the Raspberry Pi
and MapReduce to introduce students to emerging but radically different embedded
and specialized platforms~\cite{Nutt:2006:ASC}.

The rest of the paper is organized as follows. In \S\ref{sec:background}, we
present the process that was undertaken to design the course. The layout and the
progression of the course is presented in \S\ref{sec:layout}. In addition the
section also describes the hands-on component (\S\ref{sec:lab}). We dissect the
evaluation part of the course---Assignments, Projects, Code Reviews, and
Quizzes---in \S\ref{sec:evaluation}. A discussion on student feedback and
lessons learnt, and grading is presented in \S\ref{sec:discussion}.
We conclude in \S\ref{sec:conclusion}.

\section{Background}\label{sec:background}
The instructor was provided with the profiles of all students in the run-up to
the start of term. A large number of students had a computer science background,
followed by engineering and natural sciences. Their interests spanned various
fields of computer science for their theses as well as future PhD plans,
including systems, artificial intelligence, data mining, machine learning,
software engineering, and databases. One common denominator was the desire to
gain hands on experience with cutting technologies and make an impact in the
real-world beyond the classroom. Furthermore, a good fraction had used C, C++,
and/or Java in the past. Based on this diversity in background and future plans,
one of the goals of the course was to provide everyone with a level playing
field.

The choice of Python as the programming language of choice was driven by a
number of factors. The scripting nature of Python enables students to learn
programming by incrementally executing instructions in the interpreter. In
addition, its support for multiple programming paradigms, including imperative,
object oriented, functional, and event-driven is extremely useful in comparing
and contrasting these paradigms and illustrating their potential use-cases.
Furthermore, the standard library includes a large suite of packages,
encompassing diverse cases such as testing and cryptography. Moreover, the
availability of a large number of external packages for natural language
processing, machine learning, visualization, etc. makes it attractive for usage
in domain specific settings. For instance, the
SciPy\footnote{\url{http://www.scipy.org/}} library has become the tool of
choice for the scientific community as a thin wrapper for highly efficient C and
Fortran code. Finally, Python increases programmer productivity through its
dynamic typing, built in data structures, and concise syntax.

In terms of duration, the course stretched a full semester with a total of 27
lectures spanning 14 weeks. In addition, each week also had a 2 hour long
hands-on session, except the first and last week of the semester, for a total of
12 sessions.

\begin{table}[ht]
{%
\begin{tabular}{|l|l|}
\hline
\textbf{No.}   & \textbf{Title}\\\hline
1.  & \_\_init\_\_()\\\hline
2. & Control Flow\\\hline
3. & Methods and Data Structures\\\hline
4. & Object Oriented Programming\\\hline
5. & I/O\\\hline
6. & Threading\\\hline
7. & Multiprocessing\\\hline
8. & Functional Programming\\\hline
9. & Networking\\\hline
10. & Book-keeping\\\hline
11. & Security\\\hline
12. & Event-driven Programming\\\hline
13. & Graphical User Interface\\\hline
14. & Big Data and Warehouse-scale Computing\\\hline
15. & Scientific Computing\\\hline
16. & Plotting\\\hline
17. & Data Mining and Machine Learning\\\hline
18. & Image Manipulation\\\hline
19. & Natural Language Processing\\\hline
20. & Audio/Video\\\hline
21. & Graph Theory\\\hline
22. & Computer Vision\\\hline
23. & Network Emulation\\\hline
24. & Raw Packet Manipulation\\\hline
25. & Raspberry Pi\\\hline
26. & MapReduce -- Theory and implementation\\\hline
27. & MapReduce -- Applications\\\hline
\end{tabular}
\caption{Lecture Layout\label{tab:lectures}}}
\end{table}

\section{Course Layout}\label{sec:layout}
In this section, we describe the layout of the course (see
Table~\ref{tab:lectures} for a list of lectures) in detail. The first half of
the course, which comprised 13 lectures and introduced various programming
concepts and Python constructs is presented first. During this phase, students
were provided with code samples, most of which were executed live during the
lecture. The list primitive in Python was introduced from the first lecture to
enable students to use it to leverage existing data structures, such as queues,
stacks, etc. The first three lectures can be considered as a programming
refresher as they explored control flows, methods, and data structures. These
concepts were then leveraged to move to object oriented programming. I/O
constructs such as files, compression, and serialization, and concurrency were
covered in the following three lectures.
These advanced topics enabled students to be introduced to real-world issues,
such as the use of thread pools for optimum concurrency.

\subsection{First Half: Building a Foundation}
One lecture each was dedicated to functional programming and event-driven
programming. These two paradigms were included to introduce students to popular
methods readily used by industry practitioners. The use of Python also enabled
students to mix different paradigms. For instance, invoking a map function
callback in reaction to an interrupt. Python internally does not expose
event-driven programming constructs. Therefore, an external package,
Twisted\footnote{\url{https://twistedmatrix.com/trac/}} was employed to this
end. In the wake of event driven programming the design of graphical user
interfaces was covered in a lecture. A key trait of good programming is testing
and performance profiling~\cite{Chmiel:2004:DNE}. Interestingly, the author had
assumed that some of the students who were industry practitioners would be aware
of different testing methodologies, such as unit tests and regression tests.
When students were asked during the lecture about which testing methodologies
they used, most of them were clueless. A student even remarked that:
``we do not use any proper testing methods; if a piece of code works, it
works''. Finally, the near ubiquity of the Internet has enunciated the
importance of smart intrusion detection, encryption, and privacy. As a result,
the need of the hour is to have computer scientists who are formally trained in
these fields~\cite{Du:2008:SSI}. As a precursor, a lecture was dedicated to
cryptography and hashing in Python. It is noteworthy, that students were
encouraged not to make use of an IDE during the first half to enable them to
grapple with syntax problems manually and learn the hard way.

\subsection{Second Half: Venturing into the Real-world}
The second half of the course, spanning 14 lectures, employed the first half as
a foundation to tackle topics from other fields of computer science and science
in general. The flow of all 14 lectures was identical wherein, at the beginning
of the lecture, the topic is theoretically explained by using a few real-world
examples as motivation. Once the students have understood the core concepts
behind the topic, some problems from that domain are mapped onto programming
primitives from a specific Python library. At the end of each lecture, further
reading materials were also pointed out. The first lecture in this series
revolved around scientific computing and entailed the use of SciPy and
NumPy\footnote{\url{http://www.numpy.org/}}. For real-world applications,
examples from linear algebra were illustrated. A running example of the PageRank
algorithm was used to make a connection between an application in the wild and
matrices and eigenvectors. A key requirement of scientific computing is the
ability to graph and visualize difference artifacts. In light of this, on the
heels of the scientific computing lecture, a follow up lecture was dedicated to
plotting in Python using matplotlib\footnote{\url{http://matplotlib.org/}}.
The next lecture tackled machine learning and data mining. For the former, Naive
Bayes classification was used as a case study while for the latter, k-means
clustering was employed using the
Orange\footnote{\url{http://orange.biolab.si/}} library. Image processing and
manipulation was handled next using the Python Imaging
Library\footnote{\url{http://www.pythonware.com/products/pil/}}. The entire
topic was simplified by the assertion that images are just a collection of
pixels. The goal of the lecture was to describe the design of a simple graphics
editing application.

\noindent
Due to the ubiquitous nature of natural language processing thanks to its
applications in predictive text, information retrieval, and machine translation,
it was the topic of the next lecture. NLTK\footnote{\url{http://www.nltk.org/}}
was employed as the enabling package due to its rich feature set
(classification, tokenization, stemming, tagging, parsing, and semantic
reasoning) and sample corpora. The design of a spell checker and gender
classification was used to drive home some of the concepts. The latter also
acted as an illustrative example of classification algorithms covered in the
data mining and machine learning lecture. In continuation of multimedia
applications, the next lecture extended image manipulation to cover audio and
video. A combination of the \texttt{wave} module from the standard library and
pyglet\footnote{\url{http://www.pyglet.org/}} were used as enablers. A case
study of the popular Shazam application was used as a motivating example. To
this end, the instructor first loaded the application on his smart phone and
illustrated its usage through a few examples. Subsequently, the students were
prompted on whether they had ever wondered how Shazam works. Students were then
given a walk through of spectrogram based audio finger-printing to achieve such
audio matching~\cite{Wang:2003:Shazam}. This exercise made use of concepts from
the plotting (for the actual graph) and scientific computing (use of NumPy
arrays to hold audio content) lectures as well. The following lecture focussed
on graph theory and its use in such diverse applications as shortest path
calculation, web ranking, and social networks via the
NetworkX\footnote{\url{http://networkx.github.io/}} library which internally
uses matplotlib for visualization. A number of well known algorithms with their
applications were explored, including depth first and breadth first search,
connected components, shortest path, clique, topological sort, and minimum
spanning tree.

Continuing the multimedia applications strand, the twenty-second lecture went
into the details of computer vision or in simple-speak, \emph{replicating the
human ability to perceive objects}. Motivating examples included optical
character recognition, machine inspection, and automotive safety. Using
OpenCV\footnote{\url{http://opencv.org/}} a number of real-world examples around
edge detection, motion detection, facial recognition, and quantization were
illustrated. It is noteworthy that the last two examples required the use of
classification and k-means clustering from earlier lectures which is in line
with the mantra of the course to reuse existing knowledge to solve a problem.
The next two lectures focussed on computer networks applications. The first of
these tackled network emulation using software defined
networks~\cite{McKeown:2008:OEI}. To this end, the Python bindings for the
mininet emulator~\cite{Lantz:2010:NLR} for OpenFlow networks were employed.
Mininet was a natural choice as it allows production scale networks to be
emulated using a single machine. In addition, it builds upon the NetworkX
library which was already covered in the course in the graph theory lecture. In
the wake of network emulation, deeper insights into the network stack were
enabled by introducing students to raw packet manipulation using
Scapy\footnote{\url{http://www.secdev.org/projects/scapy/}}. The goal of the
exercise was to enable students to concisely write their own communication
protocols in user space without having to patch the kernel or override low-level
system calls.

The last 3 lectures revolved around programming for 2 radically different
environments: embedded systems such as Raspberry Pi~\cite{upton2012meet} and
distributed data intensive computing frameworks such as
MapReduce~\cite{Dean:2004:MSD}. With the recent interest in embedded devices due
to the rise of the Internet of Things, students should ideally be comfortable in
programming such devices. As a harbinger of this trend, students were given a
primer on programming Raspberry Pi General Purpose Input/Output (GPIO) using
Python\footnote{\url{http://sourceforge.net/projects/raspberry-gpio-python/}}.
In contrast to Raspberry Pi programming which does not require in-depth
knowledge of embedded systems concepts, the use of MapReduce requires quite a
bit of background in distributed systems. Therefore, an entire lecture was
dedicated to the architecture and theory behind it. In the last lecture of the
course, students were introduced to MapReduce application design patterns,
common application types, and a number of optimizations.
mrjob\footnote{\url{http://pythonhosted.org/mrjob/index.html}} which is a Python
wrapper around the open source Apache
Hadoop\footnote{\url{http://hadoop.apache.org/}} project was used as the
implementation with word count as a running example.

\subsection{Lab Component}\label{sec:lab}
Another key component was a weekly hands-on lab in which students had to get
their hands dirty using a given task under the supervision of the instructor and
TAs. To provide students with ample time to get comfortable with Python, the
first two labs consisted of simple list manipulation tasks with automated unit
tests. The next 4 labs incrementally built upon each other to make use of
concurrency, server design, testing and logging, cryptography, and GUI.
Specifically in lab 3, students were provided a link to a few online images and
were asked to implement a solution that leveraged a thread pool to download the
images and also incrementally display progress on the console. In the next lab,
students had to implement their own image server with internal logging for
debugging. On the heels of this lab, the fifth hands-on session revolved around
authenticating the clients and encrypting the images served by the server.
Finally, the sixth lecture completed this four lab exercise by requiring
students to implement a GUI for both the client and the server.

In lab 7, students had to implement the Hill cipher~\cite{hill1929cryptography}
using NumPy as an application of linear algebra to diverse problems. The
following session consisted of using supervised learning to filter out online
advertisements by making use of the canonical Internet Advertisements
dataset~\cite{Kushmerick:1999:LRI}. The next lab made use of Natural Language
Processing to study the stylistics of two books--of the student's choice--from
the online Gutenberg library\footnote{\url{http://www.gutenberg.org/}}. The
results of this study were visualized using matplotlib and annotated via the
Python Imaging Library. Graph theory concepts were analyzed in lab 10 through
the usage of topological sort to create a valid schedule from a course
syllabus\footnote{\url{http://www.cl.cam.ac.uk/teaching/1213/CST/CST.html}}. In
lab 11 students gained hands-on experience with network emulation by
implementing two data center topologies---a traditional 2-level tree and
DCell~\cite{Guo:2008:DSF}, a full bisection bandwidth topology---using mininet
and testing their performance. This exercise is similar in spirit to
reproducible experiments and results effected in networking
courses~\cite{Handigol:2012:RNE}. Finally, in the last lab embedded system
concepts were employed via the implementation of a traffic signal for a four
road intersection using a Raspberry Pi. For this exercise, students were
provided with a Raspberry Pi image which they mounted using
QEMU~\cite{Bellard:2005:QFP}.

Collectively, these hands-on lab sessions enabled students to put course
material in action through real-world applications. In terms of evaluation, lab
completion was a binary value. 

\section{Evaluation}\label{sec:evaluation}
The course evaluation was four-pronged: assignments, projects, code reviews,
and quizzes. We discuss each in turn in the following. It is noteworthy, that
the course did not have any formal written examination per se.
\subsection{Assignments}
All assignments and projects were designed based on the real-world programming
mantra~\cite{Stevenson:2006:DRP}. Each assignment was carefully chosen to ensure
that it was based on a real-world problem, reflected a current topic under
discussion in class, was sufficiently challenging, and provided students with
enough breathing space for creativity and innovation. Students who went the
extra mile were given bonus scores. In total, 6 assignments and 2 projects were
allocated in the term. All tasks were attempted in groups of 2 and each group
was provided with 3 get out of jail cards for the term.

In the first assignment, students were required to implement a memory efficient
sorting algorithm using a standard sort-merge tactic similar to the shuffle
phase in MapReduce-like systems. In this scheme, a large file is ingested one
chunk (of configurable size) at a time. This chunk is sorted and then dumped to
disk in the form of a spill and index file. After all chunks are sorted, the
index file is used to merge the sorted content of the spill file to result in a
globally sorted file. This assignment enabled students to write code efficient
in both space and time while tackling a ubiquitous problem from big data. The
second assignment revolved around the implementation of a simplistic web browser
with custom rendering of HTML
2.0\footnote{\url{http://www.w3.org/MarkUp/html-spec/html-spec_5.html}}.
Students had to leverage concepts from HTTP data retrieval, XML parsing, GUI
design, and concurrency to effectively implement this task. Therefore, this
assignment tested a wide range of skills. To illustrate how libraries such as
SciPy make use of native code in C and Fortran under the hood, assignment 3
involved interfacing with native code using Python. The goal of the assignment
was to implement a sorting library in Python which internally invoked an
equivalent C library\footnote{\url{http://www.yendor.com/programming/sort/}}.

The fourth assignment made use of machine learning and NLP. Specifically,
students were asked to perform Twitter sentiment analysis to work out the most
popular political parties in Pakistan. The timing of the assignment could not
have been more perfect as it was rolled out only a few weeks before the May 2013
Parliamentary Elections in Pakistan. Therefore, providing a real-world context.
The assignment had two phases: learning and online matching. The former involved
the creation of a test dataset from Twitter and sentiment learning based on the
presence of emoticons in Tweets~\cite{Alec:2009:TSC} using different
classifiers. In the second phase, a real-time portal capable of working out the
current popularity of a political party needed to be implemented. This required
skills in machine learning for classification and NLP for \emph{n}-grams in
addition to networking, GUI, and graph plotting. The next assignment revolved
around the validation of results from a recent TCP extension, dubbed
Minion~\cite{Nowlan:2012:FSP}. Minion adds unordered delivery to TCP to enable
application layer protocols which perform their own ordering to function
efficiently while looking like TCP on the wire. Students made use of Scapy,
mininet, and matplotlib to enable this. The last assignment required the
implementation of the PageRank algorithm in MapReduce with a Wikipedia
dump\footnote{\url{http://dumps.wikimedia.org/nlwiki/latest/nlwiki-latest-pages-articles.xml.bz2}}
as the test dataset.

\subsection{Projects}
Projects were similar to assignments with two additional factors: 1) while
assignments provided a blue print of the tasks to be implemented, projects were
more open ended, requiring students to do the design work as well, and 2)
projects required more implementation time. The first project was the
implementation of a simple P2P file sharing system weakly based on the
BitTorrent protocol~\cite{Pouwelse:2005:BPF}. The skillset required for
implementation included client/server design, cryptography, concurrency, and
I/O. The second project involved the reimplementation of the classic arcade game
Asteroids\footnote{\url{http://en.wikipedia.org/wiki/Asteroids_(video_game)}}
using a simplistic wrap-around 2D view. This project enunciated the use of
linear algebra in game
design\footnote{\url{http://blog.wolfire.com/2009/07/linear-algebra-for-game-developers-part-1/}}
and touched upon various concepts covered in the course.

\subsection{Code Reviews}
Students were also tasked with reviewing each other's code in two instances.
Specifically each student was sent the anonymous code from another student. The
review was expected to include both the analysis of the structure and style of
the code as well as design of a test suite.

\subsection{Quizzes}
During the course of the semester, 4 quizzes were also conducted with questions
that required: a) writing code, b) working out the output of given code, c)
pointing out errors in given code and making corrections, or d) using provided
code as a building block in a larger solution. Collectively, these distinct
types of questions tested different abilities of the students. In addition to
these 4 written (on paper) quizzes, a surprise hands-on quiz was also conducted
in the lab which entailed counting the number of smiles, eye-wear, and people in
a particular video. Students were encouraged to use existing training datasets
from the OpenCV library.

\section{Discussion}\label{sec:discussion}
The course was taken by 30 students and the final grade distribution was A+ (1),
A (3), A- (3), B+ (11), B (9), B- (1), and C+ (2). It is important to highlight
that the initial number of students was close to 60 and around half of them
bowed out during the course of the semester. This is discussed in detail below.
For the final grade, the weight of evaluation instruments was 20\% for quizzes,
54\% for assignments, and 26\% for projects. Labs were counted as bonus points
with a weight of 6\%. Therefore, the maximum score possible was 106.

During the semester, feedback was requested from students at various milestones
both verbally and in written form. One recurring theme ran through all feedback:
students were enjoying the course but believed that they were being overwhelmed
by the sheer number of concepts being covered. Some of this feedback was
incorporated on the fly into the course. For instance, the first quiz consisted
of 4 questions which needed to be solved in 30 minutes. Most students complained
that the time was not sufficient. Therefore, in subsequent quizzes, the number
of questions was reduced to 3. Similarly, the instructor initially planned on
covering artificial intelligence and robotics within the curriculum as well but
this was left out to reduce the content and resulting workload. Excessive
curriculum may account for why some of the students dropped the course in the
first few weeks. Interestingly, all of these students also discontinued the
entire graduate degree. Therefore, there might be other factors at play in their
decision. In any case, in the future it might make sense to group similar
lectures into high level concepts and reduce their details. For instance, data
mining, machine learning, NLP, and AI can be grouped into a data and knowledge
management and discovery topic which can span multiple lectures with overlapping
concepts. Similarly, image manipulation, audio/video, and computer vision can be
merged into a large multimedia umbrella.

This diversity of topics covered in the course also proved challenging for the
instructor. Each lecture required a considerable amount of background work to
cover both the theoretical and Python implementation side. For the former the
instructor had to go through academic books and online material for each topic,
such as machine learning, data mining, etc. Following this, a few illustrative
subtopics were chosen, such as supervised learning, etc. Finally, a Python
library was selected which had support for these topics and subtopics. The
challenging step was linking abstract concepts with chunks of code while
ensuring seamless consistency and incorporating real-world examples. In
addition, the instructor could only give examples of real-world scenarios based
on second-hand knowledge rather than anecdotal experience. Therefore, in
hindsight, it would have been more useful to invite industry practitioners from
each field to give an introduction to their field and share examples of their
day-to-day usage of a specific concept. This would have enhanced the motivation
behind some topics.

\section{Conclusion}\label{sec:conclusion}
In this paper we motivated the case for a graduate level course in computer
programming. To this end, we presented the design and execution of Advanced
Real-world Programming, a course which leverages Python as a vehicle to solve
real-world problems from fields as diverse as data mining and game design. The
process undertaken to design the course was described in detail along with the
lecture layout. Evaluation instruments such as assignments, quizzes, and
projects were also dissected. Finally, we shared anecdotal information from both
during the course as well as post hoc.

The design of graduate level courses has not received due attention in computer
science education research and literature. Therefore, we hope that this paper
will get the ball rolling in this direction. Moreover, the degree of detail in
this paper should make it easy for this course to be replicated, refined,
adapted, and enhanced in other settings.

{\footnotesize \bibliographystyle{acm}
\bibliography{arwp}}

\end{document}